\documentclass[manuscript,screen,nonacm]{acmart}

\makeatletter
\@twosidefalse
\@mparswitchfalse
\renewcommand\@titlefont{\fontsize{14}{16}\selectfont\bfseries} 
\makeatother

\AtBeginDocument{%
  }

\renewcommand\footnotetextcopyrightpermission[1]{} 
\begin{document}

\title[Cloning the Self for Mental Well-Being]{Cloning the Self for Mental Well-Being: A Framework for Designing Safe and Therapeutic Self-Clone Chatbots}

\author{Mehrnoosh Sadat Shirvani}
\affiliation{%
  \institution{University of British Columbia}
  \city{Vancouver}
  \state{British Columbia}
  \country{Canada}}
\email{mehrshi@cs.ubc.ca}

\author{Jackie Crowley}
\authornote{Both authors contributed equally to this research.}
\affiliation{%
  \institution{University of British Columbia}
  \city{Vancouver}
  \state{British Columbia}
  \country{Canada}}
\email{jackiec@student.ubc.ca}

\author{Cher Peng}
\authornotemark[1]
\affiliation{%
  \institution{University of British Columbia}
  \city{Vancouver}
  \state{British Columbia}
  \country{Canada}}
\email{xcher@student.ubc.ca}

\author{Jackie Liu}
\affiliation{%
  \institution{University of British Columbia}
  \city{Vancouver}
  \state{British Columbia}
  \country{Canada}}
\email{anjieliu@cs.ubc.ca}

\author{Thomas Chao}
\affiliation{%
\institution{University of British Columbia}
  \city{Vancouver}
  \state{British Columbia}
  \country{Canada}}
\email{thomas.chao@ubc.ca}

\author{Suky Martinez}
\affiliation{%
\institution{Johns Hopkins University School of Medicine}
  \city{Baltimore}
  \state{Maryland}
  \country{USA}}
\email{smart209@jh.edu}

\author{Laura Brandt}
\affiliation{%
\institution{City College of New York}
  \city{New York}
  \state{New York}
  \country{USA}}
\email{lbrandt@ccny.cuny.edu}

\author{Ig-Jae Kim}
\affiliation{%
  \institution{Korea Institute of Science and Technology}
  \city{Seoul}
  \country{Korea}
}
\email{drjay@kist.re.kr}

\author{Dongwook Yoon}
\affiliation{%
  \institution{University of British Columbia}
  \city{Vancouver}
  \state{British Columbia}
  \country{Canada}}
\email{yoon@cs.ubc.ca}

\renewcommand{\shortauthors}{Shirvani, et al.}

\begin{abstract}
\emph{\textbf{ABSTRACT}}As digital tools increasingly mediate mental health care, \emph{self-clone chatbots} can offer a uniquely novel approach to \emph{intra}-personal exploration and self-derived support. Trained to replicate users’ conversational patterns, self-clones allow users to talk to themselves through their digital replicas. Despite the promises, these systems may carry risks around identity confusion, negative reinforcement, and blurred user agency. Through interviews with 16 mental health professionals and 6 general users, we aim to uncover tensions and design opportunities in this emerging space to guide responsible self-clone design. Our analysis produces a design framework organized around three priorities: (1) defining goals and grounding the approach in existing therapeutic models, (2) design dimensions including the self-clone persona and user-clone relationship dynamics, and (3) considerations for minimizing potential emotional and ethical harms. This framework contributes an interdisciplinary foundation for designing self-clone chatbots as AI-mediated self-interaction tools that are emotionally and ethically attuned in mental health contexts.
\end{abstract}


\maketitle

\fancyfoot{} 
\renewcommand\footnotetextcopyrightpermission[1]{} 
\section{Introduction}
Mental well-being\footnote{ We use \emph{mental well-being} to refer to everyday emotional functioning, distinct from \emph{mental health,} which often implies clinical concerns \cite{gautam2024concept}.} remains an urgent and persistent global concern, and despite decades of public health efforts, a large proportion of people still cope with stress, low mood, and related difficulties \cite{fan2025global, mcgrath2023age}, which, without timely professional help, this may allow everyday distress to accumulate and escalate into more serious disorders \cite{zhang2023prevalence}. Widespread limited availability of clinical services, stigma, cost, and lack of proper psycho-education mean that much of this burden remains invisible and untreated \cite{henderson2013mental, vigo2025effective}. With the advancements in generative artificial intelligence (AI), chatbot technology represents a significant shift in therapeutic intervention, providing both immediacy and cost-effectiveness \cite{daley2020preliminary, torous2020digital}. The promise chatbots hold is in redefining the contours of how mental health support is administered, offering the potential for deeply personalized support \cite{smith2023digital}.

The recent advancements in large language models \cite{brown2020language} have turned the concept of creating digital self-clones (SCs) \cite{mcilroy2022mimetic,lee2023speculating} from pure science fiction into a tangible reality, thereby enabling \emph{the externalization of inner dialogues}, and offering a distinctive potential approach to support mental well-being. Crafted using an individual's unique data, these chatbots function as a digital twin, offering users an opportunity to converse with a replica of themselves. Talking with oneself for the externalization of internal experience has long been a cornerstone of many established therapeutic approaches aimed at fostering self-exploration and awareness by creating emotional distance. Techniques such as expressive writing through imagined letters to oneself, or the Gestalt “empty chair,” which invites individuals to dialogue with imagined aspects of themselves \cite{perls1969gestalt}, exemplify this approach. A self-clone chatbot could be a powerful extension of such practices, transforming imagined inner conversations into interactive, real-time engagements, fostering heightened self-awareness and introspection. A range of experimental and real-world applications has begun demonstrating this potential: An X artist, Michelle Huang \cite{Huang_2022} harnessed the power of GPT-3 and her childhood diaries, creating \emph{a time portal} that enabled her to converse with her \emph{inner child}. Others have created AI versions of themselves or loved ones for emotional processing on platforms like Replika or Character.ai \cite{kouros2024digital}. In academic contexts, Slater et al.’s virtual reality work on self-counseling \cite{slater2019experimental}, speculation on the role and risks of AI clones \cite{huang2025mirror, lee2023speculating}, social media clones~\cite{liu2025socialmediaclonesexploring}, therapeutic potential of future clones \cite{pataranutaporn2024future} and recent findings on self-clones' effect on enhancing therapeutic engagement \cite{shirvani2025talkingaimirrordesigning} suggest early but promising directions for such tools in enhancing mental well-being, reflection, and support.

At the same time, the idea of engaging with a digital version of oneself raises serious design, ethical, and psychological questions \cite{mcilroy2022mimetic, smith2023digital, lee2023speculating}. Without clear guidelines and safeguards, designers risk creating systems that may inadvertently reinforce negative self-schemas, normalize maladaptive beliefs, or blur boundaries between reality and simulation. Public episodes already hint at these pitfalls: After interacting with her digital replicate, Singer Grimes described the chatbot's mastery over her internal monologue as "terrifying" \cite{Grimes_2023}. Talking to oneself through clones is already being positioned as a tool for emotional support and self-reflection and explored in the recent HCI literature \cite{kim2020helping, jeon2025letters, slater2019experimental}, yet the design of these systems is advancing without shared standards. In the absence of a framework, critical decisions that affect both the safety and effectiveness of these tools---decisions that may demand professional insight---are left to designers' ad-hoc judgment, leaving the promise of therapeutic benefit vulnerable to preventable harms.

Given both the potential and the peril of self-clone technology, there is a critical need for guidance on how to navigate the interdisciplinary challenges of designing and using such systems.
Consequently, this study aims to develop a comprehensive framework to support designers in building therapeutically-relevant self-clone chatbots, clinicians in assessing and integrating them into practice, researchers in structuring future studies, and policymakers in anticipating governance needs so that these tools are optimally effective and safe. To this end, our study is guided by two primary research questions (RQs):
\begin{itemize}
    \item \textbf{RQ1:} What are the perceived therapeutic affordances, risks, and ethical implications of self-clone chatbots from the perspectives of mental health experts and prospective users?
    \item \textbf{RQ2:} How can these insights be synthesized into actionable design dimensions and safety guardrails to support responsible development?
\end{itemize}
To answer these questions, we conducted in-depth interviews with 16 mental health experts, including providers and researchers, and a complementary sample of prospective end-users (N=6). Our study focused specifically on supporting individuals experiencing mild, non-clinical symptoms, recognizing that more severe mental health challenges require purpose-built interventions and additional clinical oversight. This scope ensures that the resulting framework is grounded in early-stage self-help contexts rather than acute or crisis-oriented care.

In this work, we present \emph{a design framework for self-clone chatbots aimed at mental well-being support} as a compass to assist designers and practitioners in the creation of distinct types of self-clones that honor clinical wisdom, take advantage of AI's personalization strengths, and protect user safety. Rather than prescribing fixed solutions, our framework brings attention to three critical areas where design choices carry particular weight:
(1) therapeutic grounding and user considerations, which establishes the why and for whom by anchoring designs in established psychological approaches like parts-based and compassion-focused therapies while accounting for user readiness and attitudes; (2) design dimensions, which explores the critical how of implementation through key decision points, including the clone's persona (e.g., a mirror vs. a variant of the self), the intentional management of replication fidelity, and the degree of provider involvement; and (3) safety and ethical considerations, which provides crucial guardrails against risks such as reinforcing negative self-schemas and addresses ethical dilemmas surrounding user agency and data privacy. Collectively, these components offer a structured approach to navigate the complexities of creating supportive, rather than harmful, self-clone interactions.

This work makes three contributions to HCI and digital mental health. First, and most centrally, we present a comprehensive design framework for self-clone chatbots aimed at mental well-being support. The framework offers actionable guidance across three interrelated components: therapeutic grounding and user considerations, design dimensions, and safety and ethical considerations. Second, we contribute empirical insights from in-depth interviews with 16 mental health experts and 6 prospective users, surfacing their perceptions of the therapeutic affordances, limitations, risks, and ethical implications of self-clone chatbots---findings that ground and inform each component of the framework. Third, we advance an early, evidence-informed scholarly discussion in HCI about this emerging technology, positioning self-clone chatbots within established therapeutic traditions while charting responsible design directions for future research and practice.

\section{Related Works}

Our work is drawn from the growing literature on AI cloning, Large Language Model (LLM)-based conversational agents, and digital mental health. We synthesize insights from HCI and psychology to situate our contribution within these overlapping domains, tracing how they align with current psychological knowledge, shape emerging visions of AI-mediated care, and highlight recurring concerns around safety, trust, and ethics.

\subsection{AI Clones and LLM-based Chatbots}
    Technologies that replicate an individual’s perceptual or cognitive signature are rapidly converging into what HCI researchers now call AI clones. Early advances in synthetic media, such as photorealistic deep-fakes that transpose a person's face into new video contexts \cite{westerlund2019emergence} and neural voice-cloning systems that can synthesize the timbre of a speaker from only a few seconds of audio \cite{ruggiero2021voice}, have made it technically feasible to copy surface attributes of real people. In legal scholarship, Truby and Brown \cite{truby2021human} extended this feasibility to cognition by introducing 'digital thought clones' that replicate the underlying decision-making logic of a consumer for microtargeting. In HCI, this line of work becomes more concrete in McIlroy-Young et al.’s \cite{mcilroy2022mimetic} concept of \emph{mimetic model}, an algorithm trained on a single person’s behavioural data to simulate their future actions in a bounded domain. Synthesizing these threads, Lee et al. \cite{lee2023speculating} define AI clones as interactive agents that embody multiple facets of a real individual by fine-tuning large models on the private data of that person, while warning of risks such as identity fragmentation and unhealthy dependence.
    
    Although AI clones may be deployed by users other than the individuals they represent, such as Liu et al.’s \cite{liu2025socialmediaclonesexploring} exploration of social-media clones, a growing body of academic work centers on self-clones: agents designed to replicate aspects of a user’s own identity, history, or communicative style for direct use by the person after whom they are modeled. Academic efforts include VR self-counseling systems in which users converse with an embodied version of themselves \cite{slater2019experimental}, diary-derived reconstructions of past selves \cite{huang2025mirror}, and an interactive future-self agent \cite{pataranutaporn2024future}, which in a similar study by Shirvani et al. \cite{shirvani2025talkingaimirrordesigning} has shown promising potential to increase therapeutic engagement. These early explorations show that such personalized agents can deepen self-presence and emotional resonance, allowing people to revisit past identities, rehearse future selves, or externalize inner dialogues in ways that feel more meaningful than interactions with generic chatbots \cite{jo2024neural, park2021wrote}. Yet this same intimacy introduces distinct risks: self-clones may reinforce maladaptive self-beliefs, distort autobiographical memory, or trigger uncanny discomfort when the clone contradicts a user’s self-concept \cite{lee2023speculating, huang2025mirror}. Such systems also raise concerns about autonomy and boundary erosion, as routinely engaging a simulated self can blur distinctions between self and model, with uncertain implications for identity coherence and long-term psychological wellbeing \cite{lee2023speculating, liu2025socialmediaclonesexploring}.

    Parallel progress in LLM-based mental-health chatbots shows how such generative models can already contribute to digital mental-health interventions. Zero-shot experiments with ChatGPT-3.5 achieve high F1 scores for stress and depression detection on social media text, hinting at their value as rapid screening tools \cite{lamichhane2023evaluation}. Follow-up work confirms that GPT models can flag suicidal ideation with competitive accuracy, although fine-tuned transformers still outperform them on nuanced risk gradations \cite{ghanadian2023chatgpt}. In clinical settings, GPT-4 has been shown to predict imminent crises among telemental health clients as well as expert clinicians \cite{lee2024large, levkovich2023suicide}, and its therapeutic responses are often rated more empathic than those of licensed therapists in a Turing-style blinded test \cite{hatch2025eliza}. Comparative analyses further find that ChatGPT’s responses in help-seeking dialogues are longer, more emotionally positive, and contain richer coping suggestions than those in benchmark datasets \cite{naher2024can}. However, legacy rule-based agents like Woebot - whose scripted cognitive behavioral therapy exercises produced significant symptom reductions in a 2021 randomized controlled trial \cite{prochaska2021therapeutic} - still set the standard for safety and fidelity, as observed in earlier chatbot studies targeting symptoms such as depression, anxiety \cite{ahmed2021review}, PTSD \cite{ly2017fully}, and suicide risk \cite{wibhowo2021virtual}. This underscores the need to combine LLM flexibility with clinically verified content. Our work extends this line of research by combining the expressive flexibility of LLMs with the identity-preserving qualities of AI clones, exploring how such systems can be responsibly leveraged for mental well-being contexts.

\subsection{Landscape of Mental Health Chatbots}
In academic settings, early mental health chatbots were typically proof-of-concept systems delivering structured, evidence-based interventions—most commonly variants of cognitive behavioral therapy (CBT)—through brief, guided exchanges \cite{denecke2022implementation, fitzpatrick2017delivering}. Systems such as Woebot and Tess showed that short, high-frequency CBT-style interactions can reduce depressive and anxiety symptoms in controlled trials, establishing these agents as credible low-intensity supports \cite{prochaska2021therapeutic}. Over time, some transitioned into consumer-facing apps (e.g., Woebot, Wysa), integrating automated coaching, structured CBT exercises, and user-friendly interfaces, and are generally positioned as adjunct self-help tools that offer quick check-ins and coping strategies \cite{malik2022evaluating}. In parallel, social or companion-oriented chatbots such as Replika and Character.ai, though not explicitly built as mental health interventions, have been widely appropriated for emotional disclosure, coping with loneliness, and para-social intimacy \cite{ta2020user, skjuve2021my, lu2025utilizing, yuan2025mental}. These agents often support deeper narrative personalization and user-defined personas, but lack therapeutic framing or clinical oversight; recent reports of chatbot-related emotional dependence and suicidality highlight the risks of their unregulated use \cite{laestadius2024too, de2025emotional}.

Across these systems, several cross-cutting benefits consistently emerge: 24/7 availability, reduced stigma compared to traditional therapy, increased willingness to self-disclose, and scalable delivery of structured coping strategies \cite{haque2023overview}. Chatbots grounded in established therapeutic models tend to yield the most reliable improvements, with meta-analyses showing moderate reductions in depression and anxiety when evidence-based protocols guide conversational flow \cite{li2023systematic}. Nonetheless, personalization in current mental health chatbots remains largely shallow—typically limited to tone or exercise recommendations—rather than a deep contextual understanding of users’ histories, identities, and long-term patterns \cite{laymouna2024roles, seitz2024artificial, meadi2025exploring}. Empirical work shows that when a chatbot’s appearance or conversational style aligns with aspects of a user’s identity, users report greater trust, perceived social presence, and willingness to disclose, reflecting homophily and perceived-similarity effects \cite{chen2024effects, li2023influence, brandtzaeg2022my}. Conversely, limited continuity, contextual memory, or perceived mutual understanding can weaken trust and therapeutic alliance and reduce long-term engagement---factors known to be crucial in both human therapy and human–AI interactions \cite{shen2024empathy}. Related research on human–AI interaction shows that people tend to feel greater relational closeness and trust toward agents they perceive as similar or familiar, with similarity increasing identification and continued engagement \cite{lim2025artificial, brandtzaeg2022my}. Together, these gaps suggest significant promise for chatbots that can offer a richer sense of connection, deeper contextual understanding, and more sustained, personalized support—conditions under which self-clone chatbots may be particularly impactful.

\subsection{Self-Sourced Mental Well-Being Support}
    Digital mental health interventions (DMHI) are now operationalizing these insights by making users’ \emph{own voice} the engine of support. A systematic review of 94 apps and web programs shows that personalization --- tailoring content or tone to the individual --- is already common in DMHIs, yet most systems still rely on simple rule-based tailoring rather than true self-sourced dialogue \cite{hornstein2023personalization}. Emerging HCI work points to richer alternatives: \emph{Diarybot} lets people do expressive writing with a chatbot that listens and prompts, making disclosure “significantly less difficult” and deepening reflection compared with a static journal \cite{park2021wrote}. In virtual-reality self-counseling, users alternate between two avatars (self and counselor) to hold a \textit{literal conversation with themselves}, which yields greater perceived change than talking to a scripted therapist avatar \cite{slater2019experimental}. Even the \emph{sound of one’s own cloned voice} can bolster regulation; fMRI data show that hearing personalized self-affirmations read in your own voice recruits self-referential networks more strongly than hearing another speaker \cite{jo2024neural}. These prototypes illustrate a growing design space in which AI systems do not replace the self but \emph{amplify} it---converting private inner speech into interactive, personalized, and often more persuasive digital support. Studies have also explored self-talk–oriented chatbots designed to cultivate self-compassion \cite{lee2019caring} and reduce loneliness \cite{valtolina2021charlie}, highlighting the broader wellness potential of self-sourced interaction. We build on this literature by positioning self-clone chatbots as a possible novel tool within DMHI, showcasing the unexplored power of AI systems to transform users’ own voices and inner dialogue into more persuasive, supportive, and reflective digital interactions.

\subsection{AI Outlook in Mental Health}
    A growing body of work positions AI as a way to overcome enduring gaps in mental-health provision by delivering \emph{scalable, always-on, and personalized} support. Systematic reviews of commercially available AI–enabled apps show that most tools already provide emotion or mood tracking, personalized well-being recommendations, and conversational coaching at very low marginal cost \cite{alotaibi2023review}. Early reviews also noted that AI chatbots can enhance digital scalability and access by supporting user screening, monitoring, and even therapeutic intervention \cite{boucher2021artificially, alattas2021scale}. Large surveys with community samples likewise highlight the appeal of 24/7 availability, anonymity, and reduced stigma: nearly half of Australian respondents judged AI “beneficial” for mental-health care and emphasized its convenience and affordability \cite{cross2024use}. Qualitative analyses of user reviews for popular chatbots such as \emph{Wysa} confirm that many users value the agent’s non-judgmental stance and casual, “friend-like” conversational style, reporting relief at being able to disclose feelings without human scrutiny \cite{malik2022evaluating}. Such findings align with earlier work documenting that chatbots are often perceived as less judgmental than human therapists, encouraging openness and self-disclosure \cite{radziwill2017evaluating}. Collectively, these findings underscore AI’s perceived potential to extend preventive and self-help resources to people who would otherwise forgo or delay traditional care.

    Yet user acceptance remains cautious, with trust, privacy, and empathy emerging as recurrent barriers. In cross-sectional surveys, only a minority of participants express \emph{high} trust that AI chatbots can understand complex emotions; most exhibit moderate or “wait-and-see” attitudes, citing doubts about algorithmic empathy and fear of misdiagnosis \cite{wu2024ai}. A study of 466 adults found strong interest in AI self-help but also intense concern about data security as participants overwhelmingly demanded transparency about how sensitive mood data are stored and who can access it \cite{varghese2024public}. Literature reviews of health-care chatbots echo these worries, noting that limited crisis-handling ability, opaque decision logic, and cultural bias can erode engagement and lead to early drop-out \cite{laymouna2024roles}. Other studies note that high dropout and low long-term engagement remain persistent problems across DMHIs \cite{rudd2020digital}, often driven by unmet expectations of chatbot capability \cite{goonesekera2022cognitive, adamopoulou2020chatbots}. Users, therefore, prefer AI to \emph{augment} rather than replace human clinicians, and they call for explainable, privacy-preserving designs that foreground a “human-in-the-loop” safeguard. Design studies likewise argue that integrating periodic check-ins with human providers, or clarifying AI limitations up front, can help maintain engagement while setting appropriate expectations \cite{richards2012computer}. Addressing these trust and safety hurdles is thus pivotal to realizing AI’s promise in global digital mental health. We position our contribution as advocating for AI that augments, rather than replaces, human clinicians, and as advancing accessible tools that support a future of ethically responsible AI-mediated mental health care.

\section{Method}
We adopt an \emph{interpretivist, design-oriented} stance aimed at \emph{constructing} a practice-ready framework for the safe and effective design of self-clone chatbots for mild, non-clinical use. Our research question asks what considerations should guide the design of such systems, which is inherently forward-looking: self-clone chatbots are only beginning to emerge, and many of the most pressing issues concern how such a novel interaction is \emph{ought} to be designed safely. As there is not yet a widely used, technically mature form of self-clone technology that could serve as a stable object of study as a reference point for participants, our focus is on articulating what such systems \emph{should} be and do. As a result, participants’ inputs were necessarily speculative. In our method, we treat these anticipatory judgments as legitimate and necessary data for early-stage framework building, rather than as a flaw, following prior CHI work in which experts’ projections about emerging technologies were systematically analyzed to articulate design and governance frameworks \cite{buruk2023towards,degachi2025towards,mcveigh2019shaping}. 

To operationalize this stance, we focused our goal on \emph{framework building} that integrates empirical perspectives with extant scholarship. We therefore follow a qualitative Conceptual Framework Analysis approach \cite{jabareen2009building}, supported by reflexive thematic analysis \cite{braun2006using}. Concretely, we: (i) elicited stakeholder concepts, concerns, and design judgments through interviews; ((ii) iteratively compared these accounts with one another and with constructs from the literature, applying a logic of constant comparison inspired by grounded theory \cite{glaser2017discovery}; and (iii) articulated recurring relationships, trade-offs, and design tensions to yield a framework. While we do not claim a full grounded theory methodology, we leveraged this comparative technique to mitigate risks commonly associated with speculative inputs---such as bias, low predictive validity, and overgeneralization---by triangulating across theory, across individuals, and across participant types (experts versus prospective end-users), continually checking whether proposed considerations held beyond a single account or group. During the semi-structured interviews, we also probed for thick and specific descriptions grounded in concrete cases from participants’ clinical practice or personal experience, using hypothetical futures as an extension of, rather than a substitute for, their real-world expertise.

\subsection{Participants}
We primarily relied on mental-health experts (N=16) as our core source of insight, as their deep understanding of therapeutic processes, clinical risks, and evidence-based practice is essential for developing robust, safe, and theoretically grounded design principles for self-clone systems. However, we also recognized that an exclusively expert-driven lens could overemphasize clinical caution or introduce assumptions misaligned with end-users’ values, expectations, and goals. To balance this, we incorporated input from non-expert users as a complementary check and form of triangulation on expert assumptions, particularly on non-therapeutic, user-experience–focused areas \cite{denzin2017research}. We continued recruiting non-expert participants (N=6) until we reached thematic saturation, at which point additional interviews no longer introduced new insights in our exploratory work \cite{guest2006many}. While expert perspectives ground the framework in therapeutic rigor and safety, non-experts, who are often less constrained by clinical risk and liability, bring more openness for emotionally resonant and creative viewpoints. Their perspectives help surface interpretive gaps and opportunities that may fall outside traditional expert viewpoints. Together, these contributions support us in creating a more well-rounded and contextually grounded framework. 

Our inclusion criteria for mental health professionals required the possession of a higher degree in psychology or related fields and a minimum of two years of experience in mental health research or clinical practice. We employed the snowball sampling technique, starting from the personal social networks of the (co)authors and extending to digital channels that include social media and university newsletters. From the list of volunteers, we used purposeful sampling to optimize not only diversity but also the overall depth and breadth of experiences our experts have in the domain. With an average age of 37.37 years (SD = 10.69) and an equal distribution of women and men, the expert participants had a mean of 6.75 years (SD = 6.95) of research experience in mental health and an average of 9.12 years (SD = 7.49) of clinical experience. For details, please refer to Table \ref{table:demographics}. 

\begin{table}
\footnotesize
\centering
  \includegraphics[trim=40 100 40 50,clip,width=1.0\columnwidth]{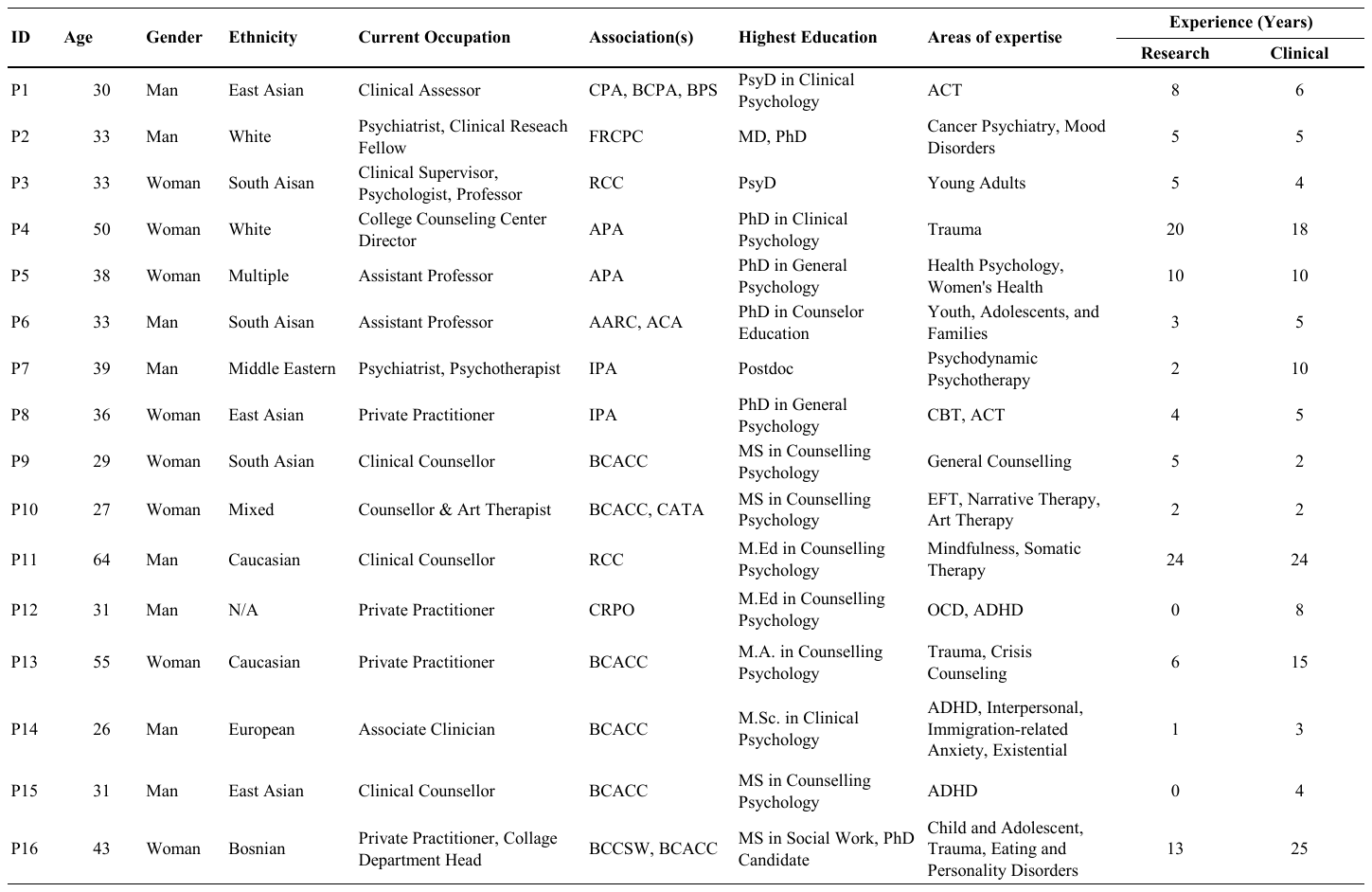}
\caption{Summary of expert participant demographics and specialization (See Appendix \ref{a} for an explanation of abbreviations)}
\label{table:demographics}
\Description{The table summarizes the demographics of expert participants (P1–P16), including age, gender, ethnicity, current occupation and affiliations, highest degree, area of expertise, and years of experience, thereby highlighting the diversity of the sample.}
\end{table}

We obtained demographic information via an online survey prior to the interview sessions. Subsequently, further data were collected, including participants' familiarity with and perspectives on existing mental health chatbots. Within our expert participants, including P1-P16, apart from one participant (P2) who had previous research experience in AI-related fields, the rest of the respondents reported limited or no prior experience with AI tools. Despite this, all participants indicated a general understanding of the concept. All the participants reported a neutral or optimistic viewpoint regarding the potential utility of such tools in the provision of mental health support.

Recruitment for the prospective, non-expert end-user group, including P17-P22, was conducted via word-of-mouth, supplemented with purposive sampling to ensure diversity in gender and experience with counseling services. Our inclusion criteria required participants to be 19 years of age or older. To ensure participant safety, we excluded individuals who reported current suicidal or homicidal ideations, as well as those with self-reported severe mental health issues accompanied by significant difficulty in coping with their symptoms. With an average age of 21.8 years (SD = 1.94) and an equal distribution of women and men, this group reported a mean of .60 years (SD = 1.10) of experience with counseling, therapy, or other mental health practices. Non-expert participants were compensated \$30 for their time.

\subsection{Data Collection}
\label{datacollection}
We utilized semi-structured interviews on Zoom, lasting around one hour, to give the flexibility to both the participants and the interviewer to expand on relevant points and views as needed. Since self-clone chatbots are a novel concept that participants were unlikely to have encountered before, we began each session by aligning on a shared definition. We introduced self-clone chatbots as interactive, text-based AI systems that mirror an individual’s way of thinking and talking (e.g., personality, language use, and problem-solving style), thereby creating an externalized digital version of that person. We emphasized that the chatbot is intended to be used only by the same person it is based on, as a tool to support mental health. 
For experts, this introduction was delivered via a brief slide presentation that included examples such as Michelle Huang’s \cite{Huang_2022} “younger self” chatbot as a real-world case, alongside higher-level scenarios developed by the research team to illustrate potential use cases of self-clone chatbots, similar to those eventually used with non-experts (Appendix~\ref{app:concepts}). To ensure comprehension before proceeding to interview questions , the interviewer then (i) asked for participants’ initial impressions of self-clone chatbots to identify any early misinterpretations or points of confusion, and (ii) invited them to describe, in their own words, what they understood a “self-clone chatbot” to be. Where misunderstandings emerged, the interviewer provided further clarification or examples until participants’ descriptions aligned with the intended framing. Our study protocol was approved by our institution’s ethics review board.

For our expert participants, we then followed a semi-structured interview guide organized around predefined discussion topics (Appendix~\ref{app:expertinterview}). Each interview began with an exploration of participants’ backgrounds and clinical or research perspectives, allowing the interviewer to identify relevant expertise, potential biases, and areas of emphasis for the conversation. This was followed by a focused discussion of the therapeutic and theoretical positioning of self-clone chatbots, and concluded with broader reflections on expected features, potential benefits, and risks. For non-expert participants, we used a similar semi-structured format but anchored the discussion in three conceptual use-case scenarios designed to make self-clone chatbots more concrete and accessible. These researcher-developed scenarios illustrated a range of potential applications and served as shared reference points throughout the interview. We opted for imagined scenarios rather than constructing bespoke self-clone chatbots for each participant so that conversations could remain speculative and unconstrained by current technical limitations, while still providing tangible examples supported by illustrative sample text exchanges approximating how an ideal self-clone chatbot might respond. All three concepts were iteratively developed and refined by the research team and are presented in Appendix~\ref{app:concepts}.

\subsection{Data Analysis}
For data analysis, we conducted a hybrid inductive–deductive thematic analysis \cite{braun2006using} informed by prevailing knowledge on chatbot design and therapeutic practice \cite{abd2019overview, casu2024ai, amershi2019guidelines, michie2013behavior}. Expert and prospective end-user interviews were analyzed separately: we first analyzed expert interviews to articulate how clinicians and researchers imagined self-clones as mental health tools and how that would translate into a design framework, then analyzed non-expert interviews to examine how prospective users’ expectations and concerns aligned with, or challenged, these expert views. Drawing on grounded theory’s constant comparative method \cite{guest2006many}, we repeatedly compared accounts within and across interviews and against relevant HCI and psychology literature, checking whether emerging considerations were supported beyond a single participant or group. While we initially expected the analysis to organize around two high-level areas—therapeutic considerations and design-relevant considerations—these deductive expectations co-evolved with inductive patterns in the data. Experts’ discussions showed that focusing on specific diagnostic labels or symptoms (e.g., anxiety) was less actionable than designing around therapeutic methods and user traits, so an initially hypothesized “problem areas and mental health symptoms” category was reframed as “therapeutic grounding.” The prominence of concerns about harm, misuse, and governance also led us to introduce “safety and ethics” as an additional high-level category. We restrict the framework to findings specific to self-clones and do not elaborate on more general design considerations for mental health AI tools.

The main analysis was led by the first author, who carried out line-by-line open coding \cite{smith2024qualitative} and generated over 160 fine-grained codes capturing concrete practices, imagined use cases, and concerns (e.g., “critical moments in therapy,” “trust building over time,” “delegating monitoring to the clone”). These codes were then iteratively clustered through axial coding into mid-level categories (e.g., “risk of reinforcing maladaptive patterns,” “configuring similarity to the self”), which were reviewed, refined, and interpreted in weekly meetings of our interdisciplinary team of HCI and psychology researchers, eventually yielding 27 conceptual categories (e.g., “SC as mirror,” “inevitability of imperfect clone”). Consistent with prior HCI work that treats the move from qualitative accounts to design guidance as an interpretive, theory-informed step rather than a direct translation \cite{dourish2006implications,lindley2020legibility}, we first developed themes grounded in interview data and then abstracted these into design considerations and framework dimensions situated within existing chatbot and HCI literature. We also conducted member checking with expert participants, inviting them to review and, if needed, clarify, adjust, or withdraw statements, which helped confirm the accuracy and fairness of our representations. Non-expert interviews were subsequently coded primarily deductively by mapping participants’ accounts onto these expert-derived categories to test the robustness, scope, and limits of the emerging framework; where their accounts did not fit cleanly, we revisited and, when needed, adjusted category boundaries. As we analyzed the non-expert interviews, participants’ accounts largely mapped onto the expert-derived categories and rarely added new insights, which we took as evidence that the core themes were saturated and robust across participant groups \cite{guest2006many}.

\section{Design Framework}
\begin{figure}[h]
  \centering
  \includegraphics[width=\linewidth]{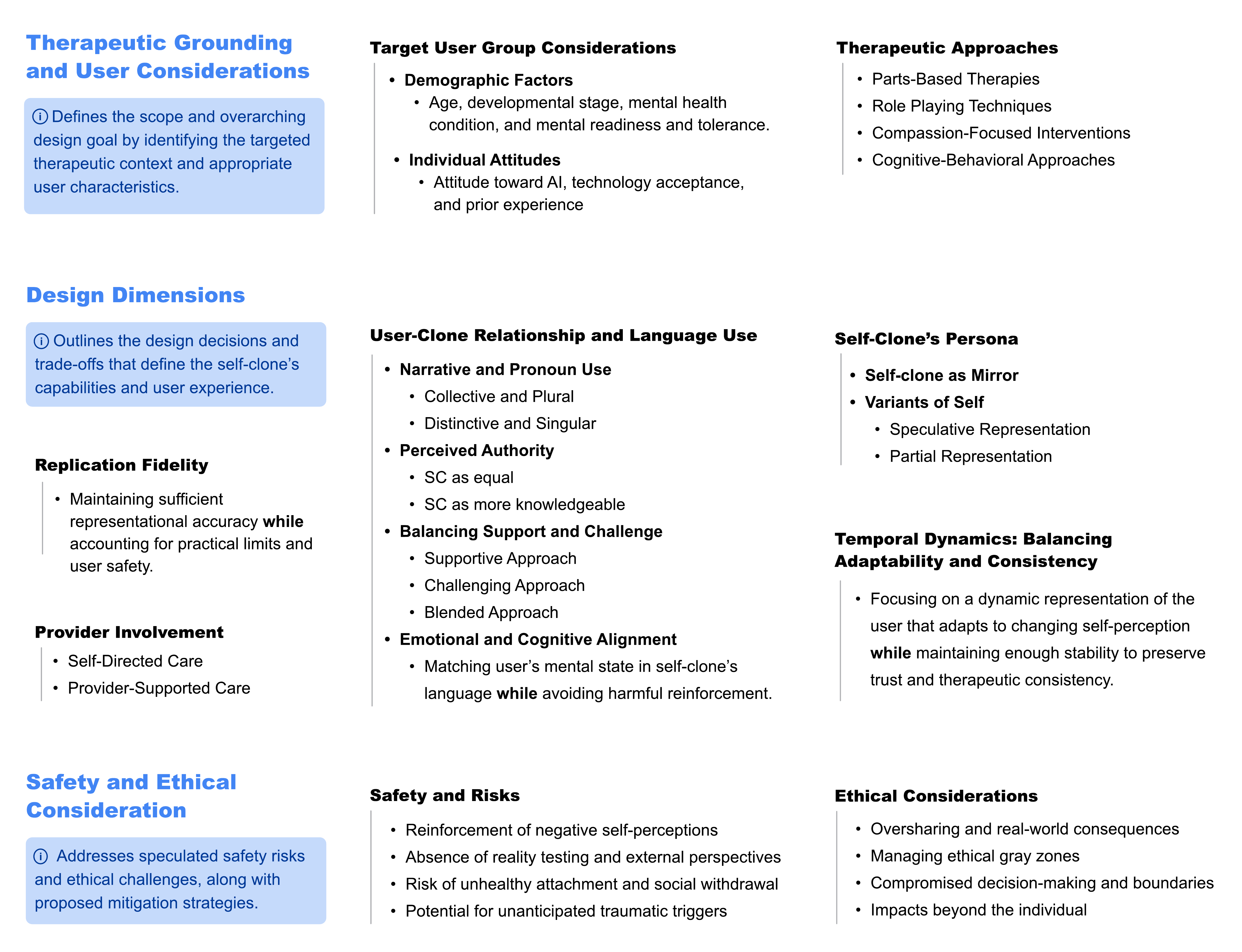}
  \caption{Overview of the key components of the proposed design framework.}
  \Description{This figure provides a summarized overview of the design framework, emphasizing high-level headings and incorporating the three levels of therapeutic grounding and user considerations, design dimensions, and safety and ethical considerations.}
\label{fig:designframework}
\end{figure}

This section presents the structured framework for designing AI self-clones aimed at supporting mental well-being. The framework comprises three interrelated components: 1) theoretical grounding and user considerations, 2) design dimensions, and 3) safety and ethics. Together, they guide designers in the responsible and effective development of self-clones by aligning user needs, credibility, and system constraints. Consistent with the recommendation of the majority of our expert participants, we position self-clones not as one-size-fits-all replacements for professional care, but as supplementary tools for individuals with mild symptoms or those seeking low-barrier entry into support; addressing clinical conditions remains the domain of specialized professional-led interventions, thus falling outside the scope of this work. A summary of the framework is shown in Figure~\ref{fig:designframework}.

\subsection{Therapeutic Grounding and User Considerations}
The following section helps establish a foundation and therapeutic direction for self-clones by exploring the evidence-based therapeutic approaches that best translate to self-clone concepts, along with the characteristics, capacities, and constraints of prospective users that may determine whether, when, and how such approaches can help. Together, these considerations clarify both the \emph{why} (therapeutic goals and intent) and the \emph{for whom} (target populations) that guide all subsequent design decisions.

\subsubsection{Therapeutic Approaches and Theories}
\label{therapeutic_approaches}
Grounding self-clones in established psychological theory can provide a credible foundation, helping them offer more effective and supportive experiences. Initially, we aimed to assess the potential of self-clones in addressing specific mental health conditions---such as symptoms of depression or anxiety---or in supporting particular therapeutic goals, including fostering self-awareness (P15)\footnote{Henceforth, the notation in the format ``(Pn)'' refers to participants who contributed to the corresponding conceptual category or quotation.}, enhancing self-compassion (P10), and promoting cognitive flexibility (P11). However, expert insights pointed to a broader conceptual shift: the value of self-clones lies less in targeting discrete symptoms and more in the \emph{therapeutic approaches they embody}. Given that most interventions can be applied across a spectrum of challenges---and that individual symptoms can be addressed through multiple modalities---similar self-clone interactions often emerged across different therapeutic targets. Consequently, our focus shifted from specific conditions to the underlying interventions that enable meaningful psychological support. Self-clones' strength lies in reimagining therapeutic frameworks to create new intervention opportunities rather than merely digitizing them. The approaches our expert participants deemed most compatible with the concept of self-clone are described below.

\emph{\textbf{Parts-Based Therapies}} 
Self-clones align naturally with parts-based therapeutic approaches, such as Internal Family Systems (IFS), which view the mind as made up of distinct internal “parts” or sub-personalities, each with its own function and voice \cite{schwartz2019internal, hunter2007client}.

\begin{quote}
    "This seems to be a tool that helps a person externalize parts… If a therapist were to work with this they could draw [out] your childhood self or your the self that holds anxiety, the self that wants to manage things…This would pair [well with] internal family systems approach, because there's a lot of talking as yourself to yourself or to parts of yourself (P10)."
\end{quote}

By focusing on embodying these specific parts, such as a critical inner voice or the wise self, self-clones enable users to engage in dynamic, conversational exploration of their internal landscape (P11, P16). This can also mirror practices in emotion-focused therapy, where clients\footnote{We use the term \emph{client} rather than \emph{patient} to reflect a more collaborative and active view of individuals seeking mental well-being support, in line with the preferences of many providers and our experts.} “talk to themselves (P10)” as a form of emotional processing. Experts further note that such dialogue can help users “see how silly the self-critic is (P12)” creating emotional distance from negative self-narratives (P11). Self-clones may also integrate complementary parts-based therapy methods, such as motivational interviewing, to support value clarification and goal-setting (P4).

\emph{\textbf{Role Playing Techniques}} 
Self-clones offer a natural fit for role-playing techniques drawn from Gestalt therapy \cite{perls1969gestalt}, particularly the empty chair method \cite{wagner2004gestalt}, which involves adopting new roles and engaging in dialogue with themselves or imagined others. Experts (P10, P12, P16) envisioned self-clones extending this practice by enabling users to converse with past selves or hypothetical versions of themselves, creating space for emotional processing and insight. As P11 noted, even informal written dialogues with imagined figures can lead to meaningful closure, suggesting that self-clone interactions could offer similar or even deeper effects. P16 further proposed a role-switching approach, where users could take on alternate perspectives, such as those of a parent or partner, while the self-clone reflects their own thoughts or behaviors back to them, helping reveal underlying relational dynamics and patterns.

\emph{\textbf{Compassion-Focused Interventions}} The introspective nature of self-clone dialogue makes it well-suited to Compassion-Focused Therapy (CFT) \cite{gilbert2009introducing}, which aims to reduce shame and self-criticism by cultivating a \emph{kinder} internal voice. As P4 noted, self-clones can function as tools for self-discovery, reflecting personal traits in a way that resembles how a therapist might “help you see aspects of yourself that you might not see.” Experts suggest that this process can unfold through interactions with a supportive clone, enabling individuals to recognize patterns of self-criticism to gradually adopt a more caring internal stance (P8). Practicing compassionate responses to self-judgment can reinforce the belief that individuals already possess the capacity for self-love (P1). For those who struggle with anxious thought patterns, self-clones may further serve as a source of reassurance, allowing users to “achieve that reassurance [they need] from themselves (P2)” rather than depending on external validation.

\emph{\textbf{Cognitive-Behavioral Approaches}} 
Cognitive-behavioral approaches, such as Cognitive Behavioral Therapy \cite{beck1979cognitive} and Dialectical Behavior Therapy \cite{dimeff2001dialectical}, can be meaningfully integrated into self-clone interactions to support cognitive restructuring and behavior change through guided self-dialogue. Users may employ their self-clone to “challenge themselves (P2)”, confront “cognitive distortions (P3)”, and “see scenarios from a new perspective (P4)”, fostering greater metacognitive awareness. From a behavioral standpoint, self-clones can assist users in developing self-care strategies (P4), recognizing unhelpful patterns, and identifying alternatives to “get [themselves] out of it” or better “manage stress and time (P2)”. These interactions may also include practical tools such as habit tracking, coping plans, or positive reinforcement. Additionally, self-clones can simulate alternative behavioral responses or personality traits, allowing users to safely explore \emph{what if} scenarios. As P16 suggested, one might “test how [their] self-clone would act with this change,” such as adopting a healthier lifestyle, helping users envision the effects of such a change before implementing new strategies in real life.

\subsubsection{Targeted User Groups Considerations}
The suitability of self-clones may vary significantly depending on the individual user, their relationship with their internal narratives, and their comfort level with engaging with such technology. While this personalized approach to self-reflection may provide valuable benefits for some users, others may find it ineffective or potentially jarring. This section presents the role of user-specific factors in shaping design opportunities and limitations, emphasizing the need to carefully consider who can use self-clones safely and effectively.

\emph{\textbf{Demographic factors.}} 
Designing self-clones around specific user demographics may offer a productive starting point. For example, adolescents might find value in using self-clones to explore identity-related questions (P4), with P15 suggesting it could help “to get a clear understanding” of one’s personal motivations and goals. Yet without adequate maturity or external guidance, younger users may experience the clone’s mirroring as “uncanny or bizarre” rather than supportive (P14). Older adults, by contrast, may bring deeper internal narratives or well-developed ‘parts’ that enrich engagement with a self-clone, though their experience may be limited by discomfort with or mistrust of digital tools (P12, P14). As P16 emphasizes, factors such as developmental stage, psychological readiness, and emotional capacity can critically shape the effectiveness of these tools. In this light, screening for pre-existing conditions and psychological readiness can help ensure that use remains within a person’s therapeutic window of tolerance. A design approach that is sensitive, adaptable, and responsive to these varying user needs will be essential in making self-clones truly accessible and supportive across diverse populations.

\emph{\textbf{Individual Attitudes.}} Self-clones, regardless of technical or design sophistication, ultimately function as AI agents role-playing the user, and their effectiveness depends on the user’s willingness to engage and sustain the illusion of a digital self for any meaningful interaction. This “willingness to engage… [depends on one’s] belief and understanding of AI (P14)” often falls outside designers’ control. In addition to users, providers’ perspectives are equally influential in the acceptance of self-clones. As P6 puts it, “tools are often only as effective in the hands of the therapist they’re deployed with,” underscoring the need to account for clinicians’ attitudes and biases, whether they are involved in designing, recommending, or deploying self-clones.

\subsection{Design Dimensions}
\label{design_dim}
In this section, we examine the critical design choices and trade-offs that shape the self-clone’s functionality and user experience. These include the agent’s persona, the fidelity of such clones, the perceived user-clone relationship, language impact, the tension between adaptability and consistency, and provider involvement.

\subsubsection{Self-Clone's Persona}
Self-clone chatbots, serving as digital identities, can adopt varying personas. The specific traits of each persona can shape users' perceptions of their self-clones, thereby broadening interaction possibilities while offering unique therapeutic potential. 

\emph{\textbf{Self-clone as Mirror.}} In their most fundamental form, self-clone chatbots function as digital mirrors, offering an externalized reflection of the user.

\begin{quote}
    "All day long, we're talking to ourselves…a lot of people actually think that their thoughts and their feelings are telling them the truth about themselves in the world…actually being able to see your thoughts or hear them externalized…that could be quite helpful to get a little bit of distance and really perspective (P11)"
\end{quote}

This concept may foster an inherent sense of familiarity with the chatbot, particularly if, as P13 speculated, it exhibits \emph{immediately recognizable shared traits} such as a similar sense of humor or shared word choices. As a result, users may find the self-clone more approachable; as P5 noted, "psychologically, we like things that are similar to ourselves." P11 further explained that individuals naturally adjust their communication style depending on their audience---speaking differently with a parent, friend, or partner---while a self-clone could offer a unique space where users can freely discuss a wide range of topics, including sensitive topics such as sex and money, without the social constraints typically present in human interactions. As P2 suggested, the use of \emph{mirror} as a metaphor could serve as a reminder for users that the self-clone’s therapeutic competence is ultimately a reflection of their own, and not a suitable substitute for professional help if needed.

Although the \emph{mirror} persona might imply that the self-clone would ideally align with the user’s personality, it does not necessarily have to be limited to reflecting the user’s perception of themselves. As P16 explained, a self-clone generated from \emph{another person's perception} of the user could serve as an informative tool for addressing relationship issues, helping individuals "put themselves in the shoes of their parents or their [partner]," and seeing themselves from another person's viewpoint.

\emph{\textbf{Variants of Self.}}
Self-clones can be intentionally designed to \emph{deviate} from the user’s current self-concept—partially replicating, modifying, or even contrasting with it—to support different kinds of self-exploration, perspective-taking, and psychological insight. However, excessive deviation can undermine the authenticity and believability of the clone as a self-representation. In this section, we distinguish between speculative and real representations of the self.

Speculative representations allow users to engage with an \emph{alternative self}, either through time-shifted versions or modified personality traits. For instance, a self-clone modeled after the user’s younger self could serve as a means to revisit past thought processes and experiences, effectively “giving a voice to the inner child” (P4). A self-clone representing an older version of the user could provide a glimpse into potential futures, enabling them to “begin with the end in mind” (P8) and reflect on “what they want to potentially keep the same or change” (P5) about their lives. Additionally, modifying specific traits—--such as anger or self-confidence---could offer an opportunity
to explore \emph{what-if} scenarios, helping users reflect on their perception of an “ideal self” (P10) and consider alternative ways of relating to their thoughts and emotions, through adjusting attributes like optimism and openness in their self-clones.

Self-clones can also \emph{partially} externalize real and distinct aspects of a user’s self-concept. This aligns with therapeutic frameworks such as Internal Family Systems and parts therapy, which emphasize engaging with different internal sub-personalities. Self-clones might embody supportive or nurturing facets of the self, reinforcing self-acceptance and resilience (P7), or reflect the user’s “inner critic and coach” (P8) or vulnerable self (P10), creating space to externalize self-doubt and foster a more compassionate stance. As P9 noted, externalizing emotional parts—such as an \emph{angry part} or a \emph{sad part}—may help users engage with and process their emotions more effectively.

Beyond embodying existing internal parts, we found that potential users see significant value in interacting with aspirational versions of themselves. Rather than seeking a perfect one-to-one replica, several non-expert participants expressed a desire for a self-clone that represents “a person that I want to become (P20)" or an “elevated version of myself (P18)" who is perhaps “smarter... but still emotionally intelligent (P18)". This indicates that a key design opportunity lies not in perfect replication but in creating a supportive role model based on the user's own values and goals, thereby using the self-clone as a tool for personal growth rather than just self-reflection.

\subsubsection{Replication Fidelity}
Although designing a self-clone requires mirroring the user’s characteristics, constructing a perfectly identical replica is neither feasible nor advisable. This inherent constraint of user-clone mismatches presents a powerful design \emph{opportunity}: rather than pursuing exact duplication, designers can strategically embrace gaps and inconsistencies to be re-framed as new insights rather than as limitations. One such approach is to introduce narrative elements, such as exploring \emph{what could have been} when absolute fidelity to a user’s past or present psyche is not achieved, or presenting a more optimistic viewpoint as the perspective of a potential future self who has grown and healed. Inevitable mismatches, if approached thoughtfully, can prompt new perspectives or encourage users to consider supports and solutions they might otherwise overlook. 

\begin{quote}
    "That's wrong...if you create a clone that's, in essence, a [perfect] clone, it must be identical to the client. So then you cloned the narcissism. You clone all of the insecurities. You clone all of the fears. You clone all of the judgments… You would copy everything over (P9)."
\end{quote}

The complex nature of human memory and self-awareness constrains accurate replication of oneself. Although this can be improved by diversifying the training data sources, in many cases, having a perfect clone may be unnecessary or even harmful. For instance, as P14 notes, “our understanding of past selves is quite skewed,” influenced by emotionally charged events, denial, or projection, such as when an individual might mislabel themselves as narcissistic after hearing a partner’s criticism (P16). Replicating such beliefs challenges the authenticity and effectiveness of self-clones as a mental health tool, calling for a \emph{selective approach} to trait replication. In long-term applications of self-clones, clients’ evolving self-concepts may necessitate a more complex and continuous approach to data collection. Such adaptation is essential to ensure that the clone reflects meaningful progress or regression and continues to resonate dynamically with the user, which we examine further in Section \ref{dimension:adapt}. Accurate data collection also presents challenges as core beliefs often reside in the subconscious and may surface only “revelation by revelation (P13)", which cannot be comprehensively captured in a single dataset. Commonly, people may also withhold personal details out of discomfort or uncertainty, resulting in incomplete or inaccurate reflections of themselves (P16), or the complexity of emotions and the dynamic nature of self-perception mean that what a user discloses at one time may not match their changing self-understanding (P15). We elaborate further on this emotional relationship aspect in ``Emotional and Cognitive Alignment'' in Section \ref{persona:relationship-language-use}. 

From a technical standpoint, any AI biases that diverge from a user’s lived experience can compromise authenticity and, thereby, trust and effectiveness; for instance, an overly positive tone may feel invalidating for users experiencing depression. Known limitations of current AI can impact the feasibility of some design scenarios. Commonly used large language models often rely on Western-centric data, which can introduce cultural and socioeconomic biases (P9, P16). For a user whose childhood language was not English, a past-self clone speaking fluent English could seem inauthentic and undermine trust (P16). Similarly, users from different cultural backgrounds may interpret experiences like racism or physical discipline in ways that deviate from western norms, yet generalized AI outputs may fail to reflect these nuances (P9, P16).

\subsubsection{User-Clone Relationship and Language Use}
\label{persona:relationship-language-use}
A perceived relationship between users and their clone is anticipated, which, despite the lack of a genuine human connection (P4), experts expect may support a healing environment akin to traditional therapy, with curiosity driving the initial stages of interaction (P13, P15). The following areas outline design considerations for fostering this perceived relationship.

\emph{\textbf{Narrative and Pronoun Use.}} Since text-based communication is the primary mode of interaction with self-clone chatbots, the perceived user-clone relationship is formed entirely within this medium. Therefore, it is essential to design language use thoughtfully. One critical consideration is the extent to which users perceive their self-clone as a distinct entity. Experts propose incorporating a collective narrative---shifting from an individual ``me versus you'' perspective to an inclusive ``we''---to foster a sense of companionship and shared experience (P2). This linguistic approach can make the chatbot feel more cooperative and less adversarial, reinforcing the notion that “we are in this together (P6)". However, some experts caution that such language use may, in certain cases, cause confusion or even raise identity-related concerns, such as "personality splitting (P7)"; therefore, use of a distinctive narrative that portrays the self-clone as a clearly separate entity may be preferable in certain cases to promote psychological distancing.

\emph{\textbf{Perceived Authority.}} The perceived authority of a self-clone, ranging from an equal counterpart to a more knowledgeable presence, can significantly shape the user’s experience. When the self-clone is framed as an equal, it could provide a more comforting environment, particularly beneficial for users navigating issues related to authority figures (P1). However, its interactions are also likely to be influenced by the user’s existing self-perception and mental state (P8). In this case, users may project their pre-existing self-views onto the self-clone, whether positive or negative, which could reinforce biases before meaningful engagement occurs. Conversely, framing the self-clone as a more authoritative figure may increase trust in its advice and guidance.

\emph{\textbf{Balancing Support and Challenge.}} The self-clone may adopt a supportive, challenging, or blended stance depending on the therapeutic context. Most scenarios benefit from a balance of support and constructive challenge, incorporating some unpredictability to sustain engagement (P6). A solely supportive or solely challenging clone is expected to be rarely preferred. However, a mismatch between the user’s emotional capacity or expected behavior and the clone’s primary approach could undermine the user’s sense of self-representation in the clone, and consequently, trust (P8). P3 further cautions that such mismatches may discourage users.

This need for a calibrated approach is reflected in the desires of non-expert, potential users, who articulated a nuanced preference for a dynamic that is both supportive and constructively challenging. One participant envisioned an ideal interaction as being “comforting and relatable like a friend, but it's also kind of... nudging you into that right direction (P19)". Another emphasized the need for the clone to “hold me accountable (P20)" for unethical actions, suggesting that pure, uncritical validation is insufficient. The consensus is that an effective self-clone should achieve a “good balance (P18)", acting as a trusted ally that can offer both affirmation and the gentle provocations necessary for growth

\emph{\textbf{Emotional and Cognitive Alignment.}}
Accurately reflecting and adapting to the user’s cognitive capacity and emotional state could encourage engagement and help build trust; however, replicating negative emotional states or limited cognitive capabilities could, in cases, be limiting or risky. Language is a highly subjective experience, as P12 explains: "It's not [just] about the words we use. It's about the meaning of the words and their connection to places and people." A believable self-clone, therefore, needs to recognize and incorporate these contextual nuances. Meeting users “where they are emotionally” (P5) can enhance validation and improve the reception of interactions, yet a persistent mismatch between the user’s state and the clone’s tone may lead to frustration or disengagement. While responsiveness to emotional cues is essential, maintaining appropriate boundaries is equally important to prevent unintended emotional escalation. Finally, aligning the chatbot’s linguistic style with the user’s intellectual and cognitive level presents an opportunity to deliver care in an accessible way, so that interactions remain understandable and supportive across different ages and educational backgrounds (P4).

\subsubsection{Temporal Dynamics: Balancing Adaptability and Consistency}
\label{dimension:adapt}
It can be challenging to design a self-clone that strikes the right balance between adapting to the user’s evolving self-perception and maintaining a coherent, predictable persona. As P13 notes, any counseling work will inevitably affect clients, especially their self-awareness and emotional maturity, and such changes should be expected to be continuously mirrored in their self-clones.

\begin{quote}
    "The system has got to keep up with that client… It's evolving… no client is stuck. The moment they're starting to talk in therapy, they're already on a track to improve…that client in session 10 isn't going to be the same client that I saw in session one. 
    Things are going to come up more just by virtue of talking about it as a clone. Is the system going to … only exist as if it's day one? (P13)"
\end{quote}

There are also implications for self-clone personas other than \emph{mirrors}. For instance, P15 explains that, over time, an individual’s perception of their “inner child” can shift substantially—from viewing it as a tantrum-throwing nuisance to recognizing it as a protective internal figure. These examples foreground a challenge of temporal dynamics: a self-clone that updates too slowly risks \emph{freezing} an outdated version of the self and slowing progression, whereas a clone that updates too quickly may overfit to transient moods and lose the stable perspective users rely on. Translating this dynamic progression into a self-clone designed for long-term support requires continuous learning from user interactions and anticipating realistic shifts as healing occurs, which is especially challenging given that the user’s internal thought processes are not always accessible as training data. In practice, this suggests that designers must make explicit choices about when and how the clone incorporates new information (e.g., after repeated patterns rather than single disclosures, through analysis of ongoing conversations, or via explicit new data-gathering) and how much weight is given to earlier versus more recent self-representations. One way to manage these changes is to emphasize the chatbot’s role-playing nature early on---for instance, positioning it as akin to a growing child, who learns to resemble the user more accurately over time, could lower expectations of perfect accuracy. Yet, such shifts arising from self-growth may not be consciously recognized by clients (P13), and they may instead notice changes in the clone’s behavior; if these are unexpected or excessive, they may be experienced as inconsistencies that undermine predictability and trust. As P10 points out, when the chatbot’s behavior and tone remain within clear, predictable boundaries, users are more likely to trust the interaction and feel in control.

\subsubsection{Provider Involvement}
A key design dimension for self-clones concerns how much involvement, if any, a mental health provider should have. This decision is shaped by therapeutic goals, user context, and safety needs, and entails weighing the benefits of autonomy and accessibility against the safeguards of clinical oversight. The two primary approaches for this dimension are presented below.

\emph{\textbf{Self-directed Care.}} 
A stand-alone self-clone can use the perceived shared understanding and values with the user to deliver supportive, self-directed care that is private, familiar, and user-controlled. By allowing users to receive support through their clone, which is a reflection of their own abilities (P2), the self-clone can foster a sense of control and self-efficacy. For discussing sensitive or private topics, as P11 notes, talking to a clone with shared moral values can diminish fear of judgment, and thus provide a 'safer space' to "speak openly ... where there's no consequence", encouraging deeper self-expression that may not have been easily possible with a therapist. While, similarly to other mental health chatbots, experts noted the accessibility and cost-effectiveness of a stand alone tool which offer "immediate support (P7)," particularly for those lacking a dependable support system (P5), as pointed out by P6, true impact of using such a tool without a mental health professional is yet to be determined and may vary on a case-by-case basis.

\emph{\textbf{Provider-Supported Care.}} 
Drawing directly from an individual’s own language, self-clone responses may carry unintended weight or influence, making provider-supported care particularly valuable in higher-risk or complex contexts, although this may reduce accessibility.

\begin{quote}
    \textit{“[Used for] day-to-day stuff [self-clone] would be a lot more widespread and more usable by everybody. If we start going deep ... often going too far with their certain thought processes, if there's no real person to really keep it in check, there could be more potential for self-harm (P1).”}
\end{quote}

Similar to the current practice with most mental health chatbots with less intensive goals, light supervision or alignment checks (e.g., reviewing transcripts or progress) from a mental health professional may be sufficient to ensure safety, but may not fully support treatment efficacy. As P15 notes, "The idea of what we think our [self-clone] says might actually differ from the type of dialogue that can lead to healing," suggesting that effective use may require a more active clinician role, providing real-time guidance or prompts during interactions, calling for a spectrum of provider support depending on therapeutic goals. As an emerging, highly personalized approach, many risks are difficult to anticipate, and experts emphasized that professional involvement enables safer and deeper exploration of the tool’s capabilities in more complex or sensitive contexts. For instance, P2 notes the potential of self-clones to "help with that exposure to trauma," which P9 cautions could be dangerous without proper supervision, as it may lead to the uncovering of repressed memories before the client is prepared. P7 further contemplates that clinicians can integrate the self-clone directly into their treatment plan as a tool for targeted introspection, particularly in approaches that encourage self-talk (e.g., IFS), thereby potentially accelerating therapeutic progress. P6, however, adds that in such applications, similar to art therapy, even with proven effectiveness, some clients and clinicians may harbor reservations about such tools, which highlights an important design consideration. Successful integration may depend on the user’s readiness to engage with their self-clone, which may require dedicated preparation, including establishing a clear mutual understanding of the clone's capabilities and limitations to mitigate potential biases.

\subsection{Safety and Ethical Considerations}

\subsubsection{Safety and Risks}
Since self-clones are inherently more personalized than generic counseling chatbots, they pose unique safety risks that may extend beyond standard AI guidelines. A prominent concern is reinforcing negative self-perceptions when users harbor harmful or “distorted views of themselves (P5)". By mirroring these traits, the self-clone may inadvertently perpetuate or “further ingrain delusions (P2)", creating what P3 describes as a “negative cycle” that is difficult to break.

\begin{quote}
    "If the clone only reflects what it's already hearing, and if the client has a very entrenched like negative sense of self and therefore they're inner talk is like terrible … it's just like reaffirming what they already believe about themselves…It could potentially send something like that and of itself could send somebody into like a depressive spiral or just feeling really emotionally dysregulated or overwhelmed (P10)."
\end{quote}

In such cases, setting proactive limits---such as “planning a way to break out of the loop (P3)"---can help prevent spiraling and encourage users to seek external interventions if necessary. A related risk lies in excessive positivity, stemming from either AI biases or a user's idealized self-perception and narcissistic tendencies, which can undermine personal responsibility and diminish critical insight (P14). Without balancing encouragement with constructive feedback, the self-clone may fail to challenge unhelpful thoughts, leaving users without “fresh perspective” (P14) or the ability to conduct meaningful “reality testing (P15)" of their beliefs. For some, self-clones may become an unhealthy attachment---akin to “social media addiction (P3)" or a “parental figure (P10)"---as users may prefer interacting with a familiar, like-minded entity, reducing their motivation to “talk to real people (P5)". In turn, they may face increased isolation (P12, P15) and lose access to genuine social or professional support, which is often vital for emotional growth. Additionally, confronting certain aspects of the self—especially past traumas—through a self-clone can trigger intense shame (P16), underscoring the need for an external safety net.

To address these issues, many experts emphasize the importance of transparency and user preparation. As P16 notes, “developing the strategy to cope if anything goes wrong” can include disclaimers, consent forms, and clear explanations of the self-clone’s capabilities and limitations (P14). By paralleling the “initial phase in therapy (P10)", users can be guided on how to handle distress, recognize the self-clone’s boundaries, and seek human help if their difficulties escalate. Such proactive measures—ranging from clarifying expectations to offering backup resources—ensure that self-clones remain supportive without displacing healthier coping strategies or professional guidance.

The necessity of such safeguards is underscored by the perspectives of potential users, who voiced strong concerns about reinforcing negative self-perceptions. Several non-expert participants feared that a self-clone might create an “echo chamber of negative thoughts (P22)" or serve to “negatively reinforc[e] ... bad trait (P19)", potentially leading a user to “spiral a bit (P19)". This convergence of expert and user viewpoints highlights the critical importance of mechanisms to challenges, rather than merely reflect, harmful cognitive patterns to ensure the tool remains a force for positive change.

\subsubsection{Ethics}
While self-clone chatbots share certain ethical concerns with other AI-driven mental health technologies, they also raise unique challenges tied to their identity-based design. Users may feel safer over-sharing sensitive information with a digital replica of themselves than with a human professional, yet this perceived comfort can mask real-world implications. This highlights the importance of confidentiality, considering that potentially "the client[s] would be coming to the chatbot in a space of vulnerability," more likely to discuss sensitive issues such as infidelity (P1).

Another key ethical dilemma is the implications of incorporating intentional discomfort as a mechanism for personal growth. In therapy, providers use clinical judgment and intuition to determine the appropriate timing and intensity of challenging conversations (P10). As P11 explains, “There are a lot of people who really need to be 'disturbed' by counseling” to break patterns of avoidance and enable change. However, AI lacks the ability to make nuanced, context-sensitive ethical decisions, creating gray zones where pushing a user too hard or too soon could be harmful (P13). Questions also arise regarding how and when a chatbot might be designed or obligated to intervene---such as reporting imminent risk like `suicidal thoughts'---without violating the autonomy of users in AI's the lack of clinical judgment and intuition (P10).

Experts also emphasize user agency (P16) and recommend that self-clones, similar to a care provider, only make suggestions, not decisions, helping users arrive at their own conclusions rather than imposing predefined outcomes (P1, P4). However, ethical concerns arise when the boundaries between the user and self-clone blur, especially if users begin attributing decisions to the chatbot rather than themselves. Clear role delineation is necessary to maintain a healthy user-clone dynamic and avoid potential dependency or confusion. Lastly, as P15 observes, a core ethical dilemma is identifying “who the chatbot serves.” A balance should be struck between ensuring individual well-being and societal considerations, such as preventing harm to others. Self-clones require a framework for these broader implications, with clear safeguards established to address ethically charged scenarios responsibly.

\section{Discussion}
\subsection{Stakeholder Implications and Practical Applications}
Our framework is designed to serve multiple stakeholders, each with distinct roles in the responsible development and deployment of self-clone chatbots. Here, we articulate how different groups can engage with and benefit from the framework's components.

\textbf{For Designers and Developers.} The framework provides actionable guidance for technology teams building self-clone systems. The design dimensions component (Section \ref{design_dim}) offers concrete decision points---persona selection, replication fidelity, language use---that can be translated into product specifications and feature roadmaps. For instance, developers can use the persona taxonomy (mirror vs. variants of self) to scope initial prototypes, while the guidance on balancing adaptability and consistency informs decisions about model updating frequency and memory architectures. We envision designers treating the framework as a checklist during early-stage ideation, returning to specific sections as implementation challenges arise. Critically, the safety considerations section provides guardrails that should be embedded into system requirements from the outset, rather than retrofitted after deployment.

\textbf{For Mental Health Practitioners.} Clinicians and counselors can use the framework to evaluate whether and how to incorporate self-clones into their practice. The therapeutic grounding component maps self-clone interactions onto established modalities (IFS, CFT, CBT), enabling practitioners to identify alignment with their existing theoretical orientation. For those considering provider-supported implementations, the framework clarifies a spectrum of involvement—from light oversight (reviewing transcripts) to active integration into treatment plans. Practitioners can also use the user considerations section to screen clients for appropriateness, assessing factors such as psychological readiness, developmental stage, and attitudes toward AI-mediated care. We recommend that clinicians unfamiliar with AI systems collaborate with technical teams to understand the limitations of the systems before recommending self-clones to clients.
For Researchers. The framework establishes a shared vocabulary and conceptual structure for future empirical work. Researchers can operationalize specific framework dimensions as independent variables in controlled studies—for example, comparing outcomes between mirror-persona and aspirational-self-persona conditions. The identified risks (negative reinforcement, unhealthy attachment) suggest hypotheses for longitudinal investigations. Additionally, the framework highlights under-explored areas, such as cultural adaptation and long-term efficacy, that warrant dedicated research programs.

\textbf{For Policy Makers and Regulators.} While our framework does not prescribe regulatory requirements, it surfaces considerations relevant to governance. The safety and ethics sections identify risks (data privacy, identity confusion, dependency) that may inform standards for mental health AI applications. Regulators can use the framework's distinction between self-directed and provider-supported care to consider tiered oversight models based on intervention intensity.

\subsection{Responsible Self-Clone Design}
In the emerging landscape of AI-powered self-clones, responsible design is the key to protecting users while ensuring their experiences are safe, meaningful, and ethically sound \cite{bates2019towards, floridi2022unified, jobin2019global}. Since a self-clone by nature is built from highly personal data, losing control of its outputs could expose a user’s private history or allow malicious \emph{reverse engineering} \cite{carlini2019secret,fredrikson2015model}. Designers, therefore, need a clear and robust data-governance plan that separates model training from model serving, applies encrypted on-device storage when feasible, and provides users with a simple way to revoke or wipe their clone \cite{regulation2016regulation}. Safety also means recognizing that some populations (e.g., youth, individuals with identity confusion, or those with a history of harm from impostor content) may require additional safeguards, such as session limits, human handoffs, or opt-out disclosures \cite{abdulai2025generative, saeidnia2024ethical}. Cultural sensitivity is another key consideration for responsible design: SCs trained on limited linguistic or cultural norms may alienate users from underrepresented backgrounds \cite{buolamwini2018gender, straw2020artificial}. Inclusive design practices and ongoing testing are vital to ensuring that diverse users feel accurately seen and supported.

The role of professional mental health professionals in administering a self-clone chatbot can be critical for ensuring safe and effective use. As our findings suggest, self-clones are best positioned as complementary tools within traditional therapeutic frameworks, rather than replacements for human care. Therapists can guide usage, monitor progress, and adapt interventions based on individual needs while also providing essential human oversight in situations where chatbot responses may fall short, such as crisis or high-risk scenarios \cite{kretzschmar2019can}. Embedding SCs within guided counseling contexts could also help mitigate ethical concerns by ensuring accountability and professional supervision \cite{lattie2022overview}. However, widespread adoption faces notable barriers. Clinicians and users alike may hold skeptical or negative views of mental health chatbots, driven by fear of AI \cite{kim2019fear}, doubts about digital efficacy \cite{borghouts2021barriers}, and lingering mental health stigma \cite{BHARADWAJ201757}. Addressing these concerns requires thoughtful tool introduction, education, and aligning with established psychological schools of thought that may be more suitable for such intervention—such as cognitivism—to frame SCs within recognized therapeutic models and evidence-based practice.

In addition to anticipating moments of acute risk, our findings highlight slower, more insidious pathways through which users may develop a specific unhealthy attachment to a self-clone. Because a self-clone is modeled on the user’s own language, preferences, and history, the relationship can feel less like using a tool and more like retreating into a version of oneself that is always available, easy to talk to, and increasingly familiar. The comforting escape this offers from interpersonal discomfort may intensify patterns of over-reliance observed in AI companions \cite{mellal2020obsolescence}, but with the added complication in cases the object of attachment is perceived as “me, but better” or “me, who understands.” Prior research shows that individuals can become dependent on digital tools in ways for which they were never designed through excessive use \cite{baumel2019objective, torous2020dropout}, suggesting that similar engagement with a self-clone could foster unintended emotional dependence or reinforce avoidance of human relationships. While current mental health chatbots attempt to mitigate this through periodic breaks, reflective pauses, and nudges back to offline coping, such strategies, while necessary, may not fully counter the pull of an always-available clone that feels immediately familiar and emotionally safe \cite{fitzpatrick2017delivering, morris2018towards}. We anticipate establishing clear expectations and continually reinforcing a realistic sense of what a self-clone can and cannot provide to remain one of the strongest preventive measures against misuse and overdependence. Expecting that interactions may sometimes go wrong, designers should also provide clear, compassionate exit strategies---not only for emergencies but for routine disengagement. For self-clone systems, this includes ways to step back when the conversation gets stuck in a reinforcing, repetitive loop, and prompts that encourage users to recalibrate and verify the validity of their assumptions in the inevitable absence of an outside perspective. Guided “soft landings” that help users integrate insights, re-situate the clone as one tool among many, and transition toward human support when needed can draw from therapy’s structured termination practices,  positioning the self-clone as a catalyst for growth rather than a preferred endpoint.

\subsection{Navigating Consent and Boundaries}
Informed consent is essential but poses unique challenges in the context of self-clones \cite{jobin2019global}. A user’s prior experience with mental health support or AI technologies can significantly shape their interaction with a self-clone, including their willingness to share or modify personal information \cite{lucas2014s, lee2023speculating}. This, in turn, affects the relevance and effectiveness of the clone, similar to what is observed in broader domains \cite{jamil2023cross}. To manage expectations and ensure ethical deployment, informed consent must be not only explicit but also understandable, regardless of a user’s AI literacy. However, a key tension arises: full transparency about the system’s AI-driven nature can sometimes reduce the immersive quality that supports engagement and reflection. We advocate for context-sensitive disclosure: presenting the clone’s artificial nature during onboarding and within user controls while allowing the ongoing conversation to feel natural unless safety concerns are triggered. Finally, it is critical to recognize that SC tools may affect more than just the individual user. Their design should also consider broader implications for family members, mental health professionals, and the societal norms they might influence.

Self-clone systems require deliberate strategies for communicating their limitations, as the true impact of the perceived user-clone relationship is still unknown, and users may overtrust personalized agents that feel unusually attuned to their own inner voice. When an AI system appears to “speak as me” and draws on the user’s past data, people may be especially prone to treating its responses as privileged self-knowledge—accurate readings of \emph{who I really am} or \emph{what I truly think}, rather than as one possible interpretation generated from patterns \cite{ho2018psychological, lucas2014s}. This makes transparent communication about system boundaries a crucial therapeutic safeguard for self-clones in particular. Such messaging should be tailored to users’ vulnerabilities and clearly convey that the clone’s outputs are generative approximations based on prior inputs, not memories, diagnoses, or authoritative insight into the self \cite{bickmore2005establishing}. In our context, this includes reminding users that the clone can amplify existing biases and blind spots just as easily as it can surface helpful reflections. Just-in-time disclosures---triggered when users ask the clone to infer deep motives or make high-stakes recommendations---can gently highlight that the response is one possible perspective, not an objective truth, recalibrating expectations without fully undermining immersion \cite{fogg2009behavior}. By foregrounding the self-clone’s status as a fallible model of the user rather than an oracle of the “true self,” these strategies can help maintain a grounded, reflective relationship with the tool while preserving the unique benefits that make self-clones compelling.

\subsection{Novelty and Cross-Domain Applicability}
The novelty of interacting with oneself can be a core part of the experience, offering a unique design opportunity to spark immediate interest and curiosity, giving users a compelling reason to start exploring. This early excitement can play a functional role in helping users experiment, uncover personal value, and establish an emotional connection with the tool. To carry this momentum forward, long-term engagement can be supported through thoughtful design strategies such as persuasive prompts, gamified reflections, or periodic progress summaries \cite{cheng2019gamification, deterding2011gamification,fogg2009behavior}. These techniques maintain relevance while reinforcing the clone’s purpose. Dynamic design can help users realign the SC with evolving needs, ensuring the interaction stays fresh and personally meaningful over time. We anticipate that the novelty of the self-clone concept will carry over into other domains. While our framework is grounded in therapeutic use, its principles provide a flexible foundation for designing self-clones in diverse contexts. Core design dimensions, such as the fidelity of the clone, the potential user–clone relationship, and the use of language, may be broadly applicable to other domains where self-clones are used, including education and social media \cite{liu2025socialmediaclonesexploring}. What may vary are the theoretical grounding and domain-specific risks, which would require tailored safeguards. As such, our framework offers a versatile foundation that can be expanded with domain-relevant extensions while preserving the core principles of responsible and effective SC design.

\section{Limitation and Future Work}
Our study offers foundational insights, but several limitations shape its scope and point to future directions. In the absence of a technically mature and established self-clone system, our approach necessarily relied on explanations and scenarios rather than participants’ direct experience with a deployed chatbot. Even with our efforts to support understanding (as explained in Section \ref{datacollection}), some participants may still have held partial or simplified mental models of self-clones, and most lacked technical expertise. Together, these factors may have limited their ability to accurately speculate about capabilities or risks. This is consistent with early, speculative framework building, and we therefore treat our findings as hypotheses about promising directions rather than exhaustive accounts of real-world use. Designing, deploying, and safely personalizing self-clone chatbots—in ways that respect the boundaries and safeguards outlined in our framework—falls outside the scope of this study and constitutes important future work, including practical trials with concrete implementations and explorations of how self-clones might be integrated with specific interventions (e.g., parts work) in clearly defined contexts.

While our recruitment was open, participants who chose to take part may have been skewed toward those already receptive to AI in mental well-being, so our findings may reflect more exploratory or optimistic perspectives than those of the broader population. We also focused intentionally on daily mental well-being rather than clinical concerns, an important distinction that was explicitly communicated and embedded in our framework. Although some experts speculated about clinical applications, systematically exploring the role of self-clone chatbots in clinical care---including how they might be integrated with structured interventions and supervision models---will require separate, dedicated studies with robust clinical oversight and safety protocols. In parallel, future work could extend our framework to richer modalities, such as voice or visual likeness, which may increase perceived presence and potential therapeutic impact but also introduce new risks (e.g., uncanny or overly personified clones) and design challenges not addressed in this study. These modalities call for their own empirical examination of psychological effects, additional system complexity, and practical value.

\section{Conclusion}
Grounded in interviews with mental-health experts and prospective users, our work examined how self-clone chatbots are perceived in terms of therapeutic affordances, risks, and ethical implications (RQ1), and how these insights can be synthesized into actionable design dimensions and safety guardrails for responsible development in non-clinical contexts (RQ2). For RQ1, participants saw self-clone chatbots as offering distinctive affordances, including externalizing internal dialogue, supporting perspective-taking (e.g., via “younger” or “future” selves), and making reflective or parts-based practices more accessible between or outside therapy. At the same time, experts and prospective users highlighted limitations and risks such as reinforcing negative self-perceptions, over-identification with a clone that mirrors unhelpful patterns, blurred boundaries between self-reflection and rumination, and concerns about privacy and data security. They also raised ethical questions about responsibility and accountability when a system is both “me” and “not me,” underscoring the need to position self-clone chatbots carefully within broader care ecosystems.

Addressing RQ2, we translated these perceptions into a design framework that organizes self-clone chatbots around concrete design dimensions and corresponding safety guardrails. Key dimensions include therapeutic grounding, the clone’s persona (e.g., inner part, reflective mirror), and the tone and nature of the user–clone relationship, each with associated trade-offs and risks. We articulate implications for stakeholder benefits, limits on the scope and intensity of interaction, and mechanisms for surfacing distress and escalation. Together, these dimensions and guardrails offer designers and practitioners practical support for making early-stage, cross-disciplinary decisions about if, when, and how to deploy self-clone chatbots in non-clinical mental health contexts. Our aim is to support futures in which self-clone technologies are developed not only for novelty or engagement, but with sustained attention to psychological safety, user agency, and long-term wellbeing.

\section{Acknowledgment}
This research was supported by the Korea Institute of Science and Technology (KIST) institutional program (26E0062). We sincerely appreciate their support, which made this work possible. Additional support was provided by the NSERC Discovery Grant and the NSERC CREATE programs.

\bibliographystyle{plainnat}
\bibliography{main}

\appendix

\section{EXPLANATION OF ABBREVIATIONS USED IN TABLE 1}
\label{a}
This appendix provides definitions, in alphabetical order, for the abbreviations used in Table 1. Summary of participant demographics and specialization.\\
1. AARC: Association for Assessment and Research in Counseling\\
2. ACA: American Counseling Association\\
3. ACT: Acceptance and Commitment Therapy\\
4. ADHD: Attention-Deficit/Hyperactivity Disorder\\
4. APA: American Psychological Association\\
5. BCACC: British Columbia Association of Clinical Counsellors\\
6. BCCSW: British Columbia College of Social Workers\\
7. BCPA: British Columbia Psychological Association\\
8. BPS: British Psychological Society\\
9. CATA: Canadian Art Therapy Association\\
10. CBT: Cognitive-Behavioural Therapy\\
11. CPA: Canadian Premier Assistance\\
12. CRPO: College of Registered Psychotherapists of Ontario\\
13. EFT: Emotion-Focused Therapy\\
14. FRCPC: Fellow of The Royal College of Physicians of Canada\\
15. IPA: Iranian Psychiatrist Association\\
16. IPsyA: Iranian Psychological Association\\
17. OCD: Obsessive-Compulsive Disorder\\
18. RCC: Registered Clinical Counsellor\\

\section{Expert Interview Questions}
\label{app:expertinterview}

The following questions were discussed with expert participants. While the interviews were semi-structured, allowing the interviewer to pose follow-up questions based on participants’ expertise and responses, all core topics were addressed as outlined below.

\textbf{Understanding Participant Background}
\begin{enumerate}
    \item As a psychiatrist/therapist/researcher, could you please introduce yourself and tell me about your background, training, and practice?
    \item Which school of psychology do you align yourself with? (e.g., structuralism, functionalism, Gestalt, behaviorism, psychoanalysis, humanism, cognitivism)
    \item Do you have any experience interacting with any AI systems?
    \begin{enumerate}
        \item What is your opinion on AI in general?
       \item If you have ever thought about AI contributing to mental health or therapy, how would you envision it?
    \end{enumerate}
\end{enumerate}

    \textbf{Understanding Participants' Approach to Clinical Practice/Research}
\begin{enumerate}
    \setcounter{enumi}{4}
    \item \textbf{[For researchers:]} Can you tell me more about your research and area of expertise?
    \item \textbf{[For clinicians:]} Can you walk me through one of your typical sessions with your clients?
    \begin{enumerate}
    \item What matters to you in building a relationship (therapeutic alliance) with the client?
    \item How do you present yourself in early sessions? What is your attitude in general? How come?
    \item How do you identify and manage critical moments in a session?
    \end{enumerate}
\end{enumerate}

\textbf{Therapeutic Consideration}
\begin{enumerate}
    \setcounter{enumi}{7}
     \item How do you think a self-clone chatbot could be used for supporting mental health?
     \begin{enumerate}
         \item In what situations would it be most helpful? (e.g., emotional distress, everyday check-in, etc.)
     \end{enumerate}
     \item Focusing on mental health problem areas, for which types of issues would the self-clone be most beneficial?
     \begin{enumerate}
         \item What topics do you think clients can talk about with the chatbot that might have therapeutic benefits?
     \end{enumerate}
     \item  Which intervention techniques would be effective to use in conjunction with self-clone chatbots? (CBT, DBT, Art, Psychodynamic, CFT, Gestalt, Holistic therapy)
     \begin{enumerate}
        \item How could a self-clone support that method?
         \item What specific intervention techniques won’t be effective w/ self-clones and why?
     \end{enumerate}
     \item  In what ways could the self-clone chatbot can help with current therapeutic practice?
     \begin{enumerate}
         \item If you had access to this as a therapist and could design for any goal you see fit, how would you incorporate it into your own practice?
         \item What is your opinion on using self-clones as a standalone tool?
     \end{enumerate}
 \end{enumerate}    
 
\textbf{Speculating on self-clone chatbots}
\begin{enumerate}
    \setcounter{enumi}{11}
     \item What are some key features (or requirements) that you expect a self-clone chatbot to have?
     \begin{enumerate}
         \item From a client’s perspective, what makes a chatbot feel convincing as a client’s digital ‘clone’?
     \end{enumerate}
     \item What do you think of the potential relationship between a client and their clone?
     \begin{enumerate}
         \item What do you think would shapes this relationship?
         \item How does this relationship compare to traditional therapeutic alliance?
     \end{enumerate}
     \item Focusing on the idea of interacting with clones (externalizing inner thoughts), what are the potential risks of using the chatbot?
     \begin{enumerate}
         \item How about when used with/without supervision and for vulnerable groups?
     \end{enumerate}
     \item What are the ethical conditions and requirements for self-clone chatbots in mental health?
     \begin{enumerate}
         \item Which decision in therapeutic practice can be made by the self-clone, and which ones should be made by the human?
     \end{enumerate}
     \item What challenges do you see in making the self-clone chatbot effective?
\end{enumerate}

\textbf{Wrap-Up}   
\begin{enumerate}
    \setcounter{enumi}{17}
     \item What other question did you expect but was not asked?
     \item Do you have further questions/comments on the topic?
\end{enumerate}

\section{Self-Clones Concepts used in Interviews with Non-Experts}
\label{app:concepts}
\begin{figure}[h]
    \centering
     \fbox{\includegraphics[width=0.8\linewidth]{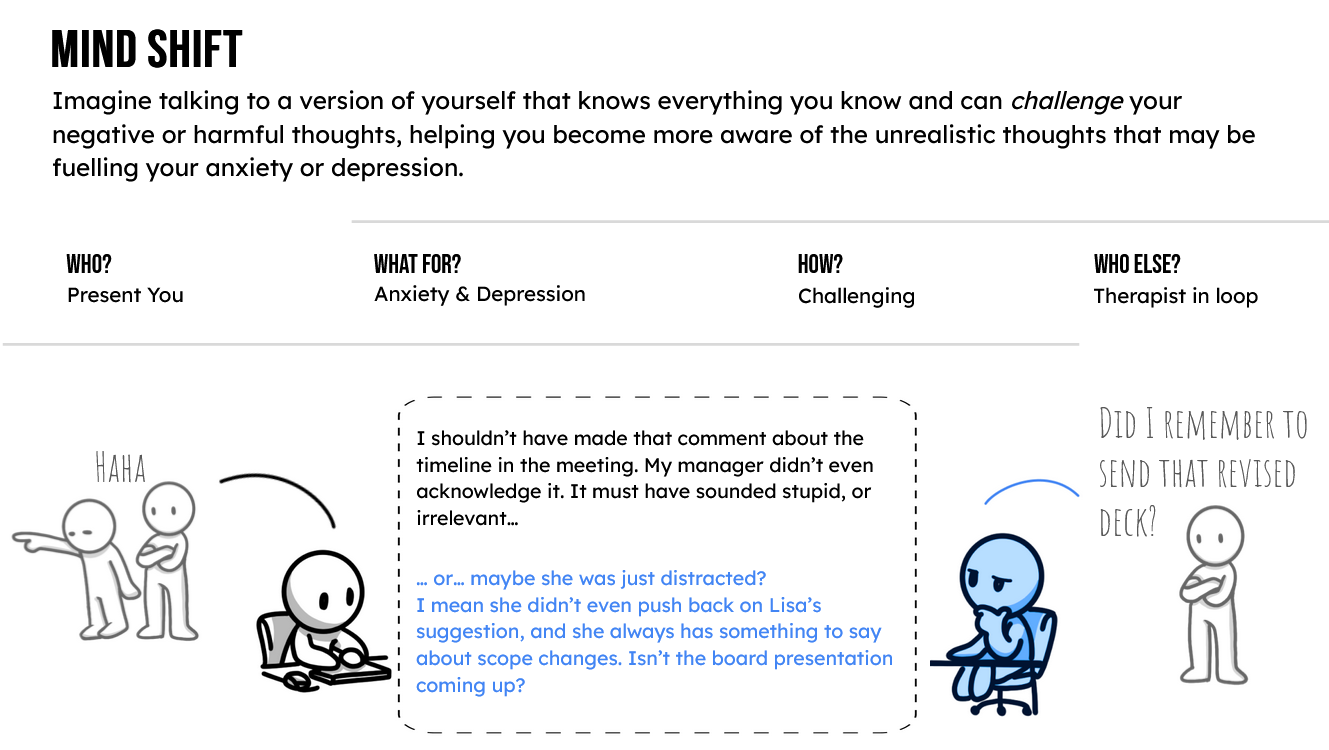}}
    \caption{Concept 1 – Mindshift: A self-clone that helps manage work-related emotions by challenging negative thoughts using personal context.}
    \Description{This figure provides a visual representation of the first concept, depicting a sample interaction between a user and their self-clone. The conversation, anchored in the present, illustrates the user’s interpretation of their manager’s response to a professional comment. The self-clone challenges this interpretation and proposes alternative explanations grounded in the user’s data, such as the manager being preoccupied with an upcoming meeting previously referenced by the user.}
    \label{fig:concept1}
\end{figure}

To support our non-expert participants in reasoning about a novel technology, we presented each concept as a visual mock-up created by the first author and iteratively refined by the research team, each including an explicit sample conversation between the user and their self-clone chatbot. In these figures, the white character represents the user and the blue character represents the self-clone, a convention explained to participants before the scenarios were introduced. The sample conversations shown in the center were designed to approximate a realistic “chat” between the user and their self-clone—both in context and visual presentation—while remaining concise enough to clearly convey the goal and outcome of each scenario within a few turns. We emphasized that these examples were illustrative rather than exhaustive, and that real-world interactions with a self-clone chatbot would likely be longer and more varied. Scenarios used are described below:

Concept 1 (Figure \ref{fig:concept1}) imagines a self-clone chatbot that “knows what you know” and helps users notice and challenge unhelpful interpretations of everyday events. The illustrative conversation is anchored in a present-day work scenario, where the user assumes their manager’s silence in a meeting means their comment was “stupid” or “irrelevant.” The self-clone, drawing on the user’s prior context (e.g., an upcoming board presentation the manager is preparing for), offers an alternative, more benign explanation—namely, that the manager may have been distracted rather than dismissive. In doing so, the self-clone explicitly challenges the user’s initial, self-critical narrative and models how a self-clone chatbot could support cognitive “mind shifts” by generating plausible, less harmful interpretations.

\begin{figure}[h]
    \centering
     \fbox{\includegraphics[width=0.8\linewidth]{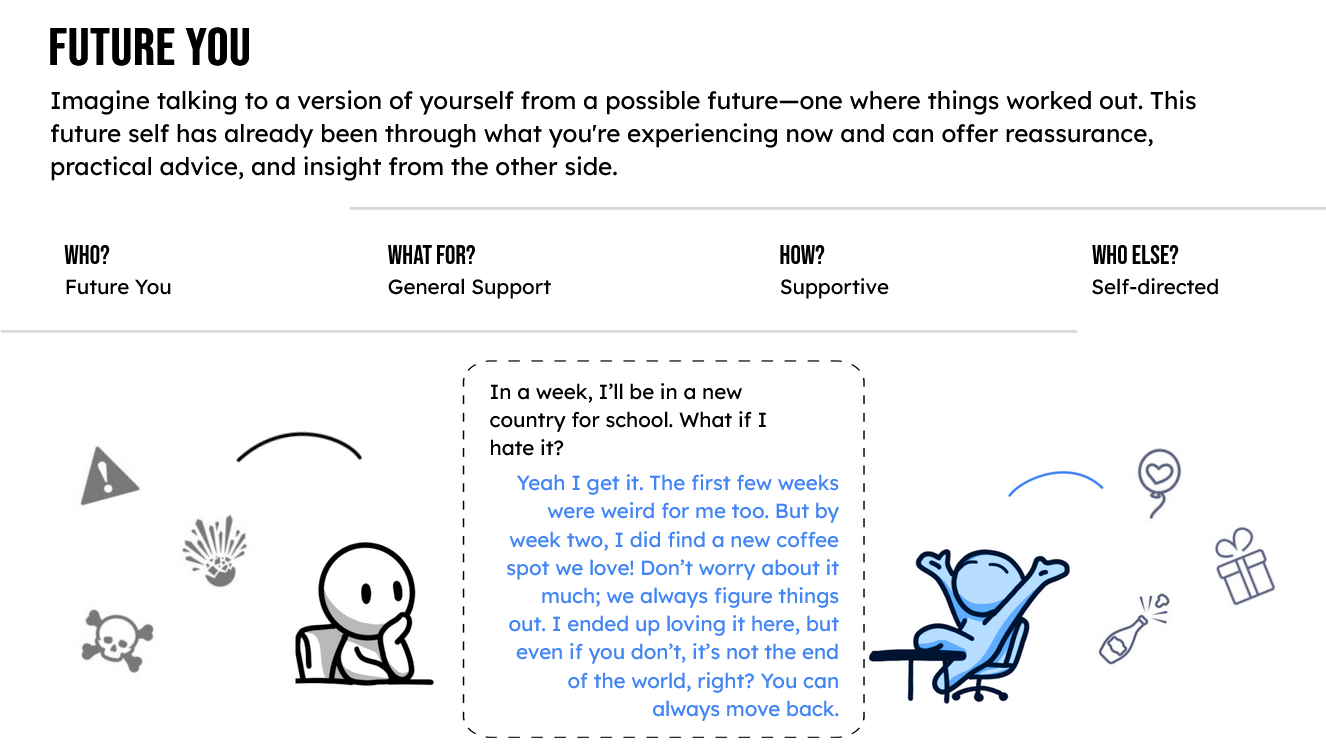}}
    \caption{Concept 2 – Future You: A self-clone representing a hopeful future self offering reassurance and perspective based on possible positive outcomes.}
    \Description{This figure presents a visual representation of the second concept, illustrating a sample interaction between a user and their future self-clone. The exchange adopts a supportive approach, addressing and alleviating the user’s concerns about relocating abroad.}
    \label{fig:concept2}
\end{figure}

Concept 2 (Figure \ref{fig:concept2}) imagines a self-clone chatbot that speaks from a hopeful near future in which the user’s current challenge has worked out, and is already attuned to the user’s typical emotional support needs. Drawing on the user’s past patterns---such as finding it helpful when worries are acknowledged, balanced with concrete positives, and paired with a “way out” for worst-case scenarios---the future self-clone offers reassurance, perspective, and specific examples of things that went well. In the illustrative conversation, the user worries about relocating abroad for school and “hating it.” The future self-clone normalizes the anxiety (“the first few weeks were weird”), highlights a small but meaningful positive (discovering a favorite coffee spot), and reminds the user that even if things do not work out, they can always move back. The scenario demonstrates how a future-oriented self-clone chatbot, tuned to the user’s established coping preferences, could provide emotionally calibrated support rather than generic reassurance.

\begin{figure}[h]
    \centering
     \fbox{\includegraphics[width=0.8\linewidth]{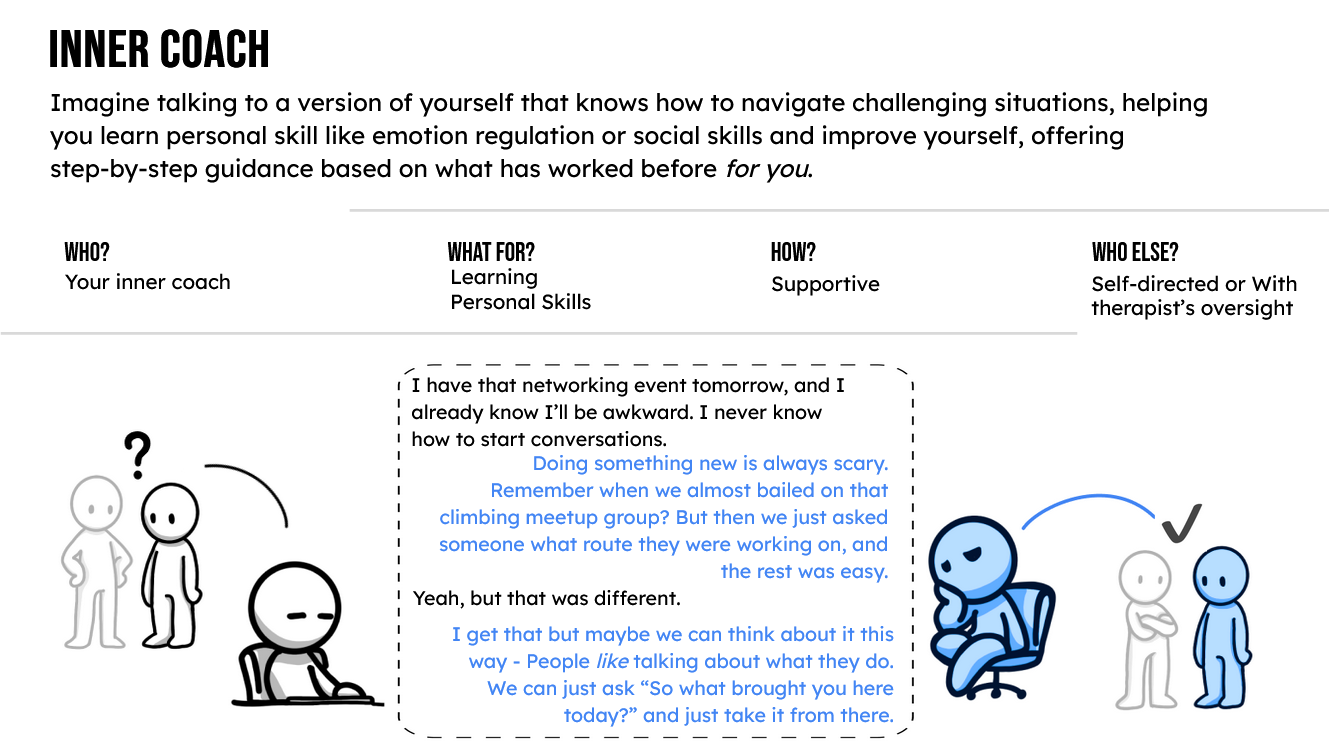}}
    \caption{Concept 3 – Inner Coach: A self-clone acting as a personalized coach, guiding self-improvement through tailored, experience-based support.}
    \Description{This figure presents a visual representation of the third concept, depicting the self-clone in the role of an inner coach. It supports the user in skill-building for a networking event by providing guidance, referencing prior experiences, and highlighting relevant similarities to reduce stress.}
    \label{fig:concept3}
\end{figure}

Our last concept (Figure \ref{fig:concept3}) imagines a self-clone chatbot acting as a personalized coach that supports self-improvement (e.g., emotion regulation, social skills) by drawing on the user’s recurring patterns and prior experiences. In the illustrative scenario, the user expresses anxiety about an upcoming networking event and doubts their ability to start conversations. The self-clone recognizes this familiar pattern, briefly validates the anxiety, and recalls a similar past situation that turned out well—offering both reassurance and a concrete example the user can draw on when considering similar solutions. When the user shows resistance, the self-clone acknowledges it and gently reframes the challenge, offering a specific, tailored strategy. The scenario illustrates how a self-clone chatbot, attuned to the user’s history and growth goals, can function as an “inner coach” that combines emotional validation with step-by-step, experience-based guidance.
\end{document}